\documentclass[twocolumn,showpacs,preprintnumbers,amsmath,amssymb,aps]{revtex4}
\usepackage{epsf}
\usepackage{color}
\usepackage{dcolumn}% Align table columns on decimal point
\usepackage{bm}% bold math
\everymath{\displaystyle}
\begin{document}
%%%%%%%%%%%%%%%%%%%%%%%%%%%%%%%%%%%%%%%%%%%%%%%%%%%%%%%%%%%%%%%%%%%%
% End of preamble and beginning of text.
%\pagestyle{empty}
%%%%%%%%%%%%%%%%%%%%%%%%%%%%%%%%%%%%%%%%%%%%%%%%%%%%%%%%%%%%%

\title{Semiclassical limit for Dirac particles interacting with a gravitational
field}

\author{\firstname{Alexander J.}~\surname{Silenko}}
\email{silenko@inp.minsk.by}
%\noaffiliation
\affiliation{Institute of Nuclear Problems, Belarusian State
University, Minsk 220080, Belarus}

\author{\firstname{Oleg V.}~\surname{Teryaev}}
\email{teryaev@thsun1.jinr.ru}
\affiliation{Bogoliubov Laboratory of Theoretical
Physics, Joint Institute for Nuclear Research,
Dubna 141980, Russia}

\date{\today}

\begin {abstract}

The behavior of spin$-1/2$ particle in a weak static gravitational
field is considered. The Dirac Hamiltonian is diagonalized by
the Foldy-Wouthuysen transformation providing also the simple form for
the momentum and spin polarization operators. The operator
equations of momentum and spin motion are derived for a first
time. Their semiclassical limit is analyzed.
The dipole spin-gravity
coupling in the previously found (another) Hamiltonian does not lead to any observable effects.
The general agreement between the quantum and classical analysis is
established, contrary to several recent claims. The expression for
gravitational Stern-Gerlach force is derived.
The helicity evolution
in the gravitational field and
corresponding accelerated frame coincides, being
the manifestation of the equivalence principle.

\end{abstract}

\pacs {04.20.Cv, 03.65.Ta, 04.25.Nx}
\maketitle

\section {Introduction}

Interaction of elementary
particles with gravitational fields poses an interesting problem
with important astrophysical applications. One of the approaches
to this problem is provided by a corresponding Dirac equation in an
external gravitational field. It was recently solved \cite{Ob1}
using the exact diagonalization by an appropriate unitary
transformation for the wide class of static gravitational fields.
However, the presence of a dipole spin-gravity
coupling in final results of Refs. \cite{Ob1,Ob2} is controversial
\cite{Nic,Ob3}. For accelerated frames, there is
not any similar coupling (see \cite{HN,Huang,HuangNi}).

There is also a related problem of disagreement between the classical formula
for the angle of particle deflection by a gravitational field and the
corresponding expression for Dirac particles
claimed recently by another author \cite{RAINBOW}.

In the present article we resolve these contradictions. The diagonalization of
Dirac equation is still insufficient to get the
semiclassical equations of spin motion formerly obtained in \cite{PK,Obzor}.
The problem is that the derivation of the equations of motion requires also the
knowledge of respective dynamical
operators, in particular, that of momentum and spin.
We investigate this problem and show that these operators
have a rather complicated form in the representation used in \cite{Ob1,Ob2},
which is because that representation, although diagonal, does not possess all
the
properties
of the Foldy-Wouthuysen (FW) one. As a result, the dipole spin-gravity
coupling appearing in \cite{Ob1,Ob2} does not lead to new observable effects.

To bypass this difficulty, we construct
%instead
the ``standard" FW representation  where the dynamical operators take the simple
form.
We derive (for the first time, up to our knowledge)
the operator equations of momentum and spin motion in a
weak spherically symmetric gravitational field and uniformly accelerated frame.
We study the
semiclassical limit of these equations to get the momentum, spin polarization
and helicity evolution.
The results fully agree with the classical gravity
(so that the disagreement found in \cite{RAINBOW} is not confirmed)
and contain quantum corrections.
In particular, the expression for the gravitational Stern-Gerlach (SG) force
acting on relativistic particles is
found.

\section {Dirac equation for particles in a static gravitational field}
%Foldy-Wouthuysen and Eriksen-Korlsrud representations}

An interaction of a spin-1/2 particle with a gravitational field is
described
by the covariant Dirac equation:
%(see Refs. \cite{Ob1,Ob2} and references
%therein).
\begin{equation}
(i\gamma^\alpha D_\alpha-m)\psi=0, ~~~~~~~\alpha =0,1,2,3,
\label{eqin}\end{equation} where $\gamma^\alpha$ are the Dirac
matrices. The system of units $\hbar=c=1$ is used. The spinor
covariant derivatives are defined by
\begin{equation}
D_\alpha=h_\alpha^iD_i, ~~~ D_i=\partial _i+\frac
i4\sigma_{\alpha\beta}\Gamma_i^{\alpha\beta},
\label{eqin2}\end{equation} where $h_\alpha^i$ and
$\Gamma_i^{\alpha\beta}=-\Gamma_i^{\beta\alpha}$ are the coframe and Lorentz
connection coefficients,
$\sigma^{\alpha\beta}=i(\gamma^\alpha\gamma^\beta-\gamma^\beta\gamma^\alpha)/2$
(see Refs. \cite{Ob1,Ob2} and references therein).
%We use the
%Greek alphabet for the indices which label the components with
%respect to a local Lorentz frame $e_\alpha =
%h_\alpha^i\partial_i$, whereas the Latin indices refer to the
%local spacetime coordinates $x^i$.
%For the Dirac matrices, the
%conventions of \cite{BD} are used.
Following these Refs. we limit ourselves to the case of the static spacetime
\begin{equation}
ds^2 = V^2\,(dx^0)^2 -
W^2\,(d\bm{r}\cdot d\bm{r}).\label{eq1}
\end{equation}
Here $V, W$ are arbitrary functions of $\bm{r}$. Particular cases
belonging to this family are pointed out in \cite{Ob1,Ob2} and
include

(i) the flat Minkowski spacetime in
accelerated frame
\begin{equation}
V = 1 + \bm{a}\cdot\bm{r}, ~~~ W = 1\label{acf}\end{equation}

and (ii) Schwarzschild spacetime in the isotropic coordinates
\begin{equation}
V =\! \left(1 - {\frac {GM}{2r}}\right)\left(1 + {\frac {GM} {2r}}
\right)^{-1}\!, ~~~ W =\! \left(1 + {\frac {GM}{2r}}\right)^2\!
\label{Sh}\end{equation} with $r = |\bm{r}|$. For metric
(\ref{eq1}), the Dirac equation can be brought to the Hamilton
form \cite{Ob1,Ob2}
\begin{equation}
i\frac {\partial \psi}{\partial t}={\cal H}\psi, ~~~
{\cal H} = \beta m V+\frac12\{{\cal F},\bm\alpha\cdot\bm p\},
\label{eq2}\end{equation}
where ${\cal F} = V/W$ and $\{\dots,\dots\}$ denotes the  anticommutator.
This equation is the starting point of our analysis.

\section {Connection between the Foldy-Wouthuysen and Eriksen-Korlsrud
representations}

The FW transformation \cite{FW} provides the
correct physical interpretation of Dirac Hamiltonians.
The important advantage of the FW representation is the
simple form \cite{FG} of polarization operator $\bm O_{FW}$ being equal to
the matrix
\begin{equation}
\bm O_{FW}= \bm \Pi = \beta \bm \Sigma. \label{fg}
\end{equation}
In principle, this form of polarization operator may be considered as
a definition of the FW representation.

In Refs. \cite{Ob1,Ob2}, an exact block-diagonalization of Hamiltonian
(\ref{eq2})
by the Eriksen-Korlsrud (EK) method \cite{EK} has been performed.
However, a block-diagonalization
%(diagonalization in two spinors)
of Hamiltonian may be nonequivalent to the FW transformation.
There exists an infinite set of representations where all the
operators are block-diagonal. Therefore, the equivalence of any
representation to the FW one should be verified.
For example, the transformation performed in Ref. \cite{Ts} for particles in a
uniform magnetic field has led to
a block-diagonal Hamiltonian. However, this Hamiltonian differs from the
corresponding Hamiltonian in the FW representation
\cite{YPhys}.

It is easy to prove the FW and EK representations are not equivalent even for
free particles. The unitary operator of
transformation from the Dirac representation to the EK one is given by
\cite{Ob1,Ob2,EK}
\begin{equation} \begin{array}{c}
U_{D\rightarrow EK}=\frac12(1+\beta J)(1+J\Lambda), ~~~ J=i\gamma_5\beta, \\
\Lambda=\frac{{\cal H}}{\sqrt{{\cal H}^2}}=\frac{\bm\alpha\cdot\bm p+\beta
m}{\epsilon},
~~~  \epsilon =\sqrt{m^2+\bm p^2}. \end{array}\label{eq3}\end{equation}

The unitary operator of transformation from the Dirac
representation to the FW one is equal to \cite{FW,JMP}
$$ U_{D\rightarrow FW}=\frac{\epsilon +m+\beta\bm\alpha\cdot\bm
p}{\sqrt{2\epsilon (\epsilon +m)}}.
$$

Therefore, the operator providing the transformation from the FW representation
to the EK one is
\begin{equation}
U_{FW\rightarrow EK}=U_{EK}U_{FW}^{-1}=
\frac{\epsilon +m+i(\bm\Pi\cdot\bm p)}{\sqrt{2\epsilon (\epsilon +m)}}.
\label{eq4}\end{equation}

For free particles, this operator does not change the form of the
Hamiltonian. However, operator (\ref{eq4}) is not equal to the
unit matrix and therefore changes the wave eigenfunctions.
Consequently, the FW and EK representations are nonequivalent.

It is easy to see that the polarization operator in the EK representation is
very different from the corresponding operator
in the FW representation even for free particles:
\begin{eqnarray}
\bm O_{EK}=U_{FW\rightarrow EK}\bm\Pi U_{FW\rightarrow EK}^{-1}
\nonumber\\
=\bm\Pi+\frac{\bm p\times\bm\Sigma}{\epsilon }+\frac{\bm p\times(\bm
p\times\bm\Pi)}
{\epsilon (\epsilon +m)}.
\label{eq5}\end{eqnarray}

For particles in external fields, this circumstance brings a
difference between Hamiltonians, especially for the terms
proportional to the polarization operator. Thus, the
block-diagonalization of the Hamiltonian needs to be fulfilled
carefully.

It is important that the forms of the position operator in two
representations also differ. In the FW representation, this
operator is just the radius vector $\bm r$ \cite{NW,CMcK}. In the
EK representation, it is given by
$$\begin{array}{c}
\bm r_{EK}=U_{FW\rightarrow EK}\bm r U_{FW\rightarrow EK}^{-1}
\\
=\bm r+\frac{\bm p\times\bm\Sigma}{2\epsilon(\epsilon
+m)}+\frac{\bm\Pi} {2\epsilon}- \frac{\bm p\cdot(\bm
p\cdot\bm\Pi)} {2\epsilon^2 (\epsilon +m)}.
\end{array}$$

Thus, the EK transformation does not lead to the FW representation.

\section {Foldy-Wouthuysen transformation for spin-1/2 particles in a static
gravitational field}

Let us transform Hamiltonian (\ref{eq2}) to the FW representation. For this
purpose, we apply the method of relativistic FW
transformation elaborated in Ref. \cite{JMP}. The validity of this method is
confirmed by the consistency of results
obtained by different methods for the electromagnetic interaction of particles
(see Ref. \cite{JMP}).
So, we expect it to be valid and provide the simple expression for dynamical
operators
also for the gravitational field.

Unfortunately,
we are unable to perform the exact FW transformation. Therefore,
we use the weak-field approximation which makes it possible to obtain the FW
Hamiltonian as a power series in parameters of an external field.
In our case this requires that
$|V-1|~,|W-1|\ll 1.$
%It is of no importance that corrections are proportional to the
%particle mass $m$. The small
%parameter is the ratio of "the energy of gravitational interaction``
%to the total energy of particle including the rest energy.

Hamilton operator (\ref{eq2}) can be written in the form
$$ {\cal H}=\beta m+{\cal E}+{\cal
O},~~~\beta{\cal E}={\cal E}\beta, ~~~\beta{\cal O}=-{\cal O}\beta, $$
where
$${\cal E}=\beta m(V-1), ~~~ {\cal O}=\frac12\{{\cal F},\bm\alpha \cdot\bm p\}
$$ mean terms commuting and anticommuting with the matrix $\beta$, respectively.

Other notations
$\bm\phi=\nabla V, ~~~ \bm f=\nabla {\cal F}~~$  follow Refs. \cite{Ob1,Ob2}.

Let us perform the FW transformation for relativistic particles with an
allowance for first-order terms in the metric tensor and its
derivatives up to the second order.

After the first transformation with the operator (see Ref. \cite{JMP})
$$U=\frac{\epsilon'+m+\beta{\cal O}}{\sqrt{2\epsilon'(\epsilon'+m)}}, ~~~
\epsilon'=\sqrt{m^2+{\cal O}^2},$$
the Hamilton operator takes the form:
$$ {\cal H}=\beta\epsilon' +{\cal E}'+{\cal O}',~~~
\beta{\cal E}'={\cal E}'\beta, ~~~\beta{\cal O}'=-{\cal O}'\beta, $$
where
\begin{widetext}
$$ \begin{array}{c}
\epsilon'=\sqrt{m^2+{\cal O}^2}=\sqrt{m^2+\bm p^2+\{\bm p^2,{\cal
F}-1\}+\frac{1}{2}
\left[\bm{\Sigma}\cdot(\bm f\times\bm p)-\bm{\Sigma}\cdot(\bm p\times\bm f)+
\nabla \cdot\bm f\right]},\\
{\cal
E}'=\frac{\beta}{2}\left\{\frac{m^2}{\epsilon },V-1\right\}-\frac{\beta
m}{4\epsilon (\epsilon +m)}\left[\bm{\Sigma}\cdot(\bm\phi\times\bm p)-
\bm{\Sigma}\cdot(\bm p\times\bm\phi)+\nabla \cdot\bm\phi\right]   \\
+\frac{\beta}{8}\cdot\frac{(2\epsilon ^3+2\epsilon ^2m+2\epsilon
m^2+m^3)m}{\epsilon ^5
(\epsilon +m)^2}(\bm p\cdot\nabla)(\bm p\cdot\bm\phi), ~~~ \epsilon =
\sqrt{m^2+\bm p^2}.
\end{array}$$
\end{widetext}

We neglect the noncommutativity of operators in small terms %that are
proportional to derivatives of the metric tensor.
The calculation of ${\cal O}'$ is unnecessary because its contribution to the
final FW Hamiltonian is of order of $(V-1)^2$.

The quantity $\epsilon'$ can be represented as
\begin{eqnarray}
\epsilon'
=\epsilon +\frac{1}{2}\left\{\frac{\bm p^2}{\epsilon },{\cal F}-1\right\}+T
\nonumber\\
+\frac{1}{4\epsilon }\left[\bm{\Sigma}\cdot(\bm f\times\bm p)-
\bm{\Sigma}\cdot(\bm p\times\bm f)+\nabla \cdot\bm f\right].
\label{eq6}\end{eqnarray}
To determine the operator $T$, it is necessary to square both parts of Eq.
(\ref{eq6}).
% and use the following operator equality ($[A,C]=0$):
% $$ABC+CBA=\{AC,B\}-[C,[A,B]]. $$
As a result of calculation,
$$ T=-\frac{\epsilon ^2+m^2}{4\epsilon ^5}(\bm p\cdot\nabla)(\bm p\cdot\bm f),$$
and the final expression for the FW Hamiltonian takes the form
\begin{widetext}
\begin{eqnarray}
{\cal H}_{FW}=\beta\epsilon +\frac{\beta}{2}\left\{\frac{m^2}{\epsilon
},V-1\right\}
+\frac{\beta}{2}\left\{\frac{\bm
p^2}{\epsilon },{\cal F}-1\right\}-\frac{\beta m}{4
\epsilon (\epsilon +m)}\biggl[\bm{\Sigma}\cdot(\bm\phi\times\bm p)-
\bm{\Sigma}\cdot(\bm p\times\bm\phi)+ \nabla\! \cdot\!\bm\phi\biggr]
\nonumber\\
+\frac{\beta m(2\epsilon ^3+2\epsilon ^2m+2\epsilon
m^2+m^3)}{8\epsilon ^5 (\epsilon +m)^2}(\bm p\cdot\!\nabla)(\bm
p\cdot\!\bm\phi)+ \frac{\beta}{4\epsilon }\left[\bm{\Sigma}\cdot(\bm
f\times\bm p)- \bm{\Sigma}\cdot(\bm p\times\bm f)+\nabla\! \cdot\!\bm
f\right]-\frac{\beta(\epsilon ^2+m^2)}{4\epsilon
^5}(\bm p\cdot\!\nabla)(\bm p\cdot\!\bm f).
\label{eq7}\end{eqnarray}
\end{widetext}

It is obvious that this expression differs from the corresponding
one derived in Refs. \cite{Ob1,Ob2}. To perform a more detailed
analysis, we can rewrite Eq. (13) from Ref. \cite{Ob2} in the
weak-field approximation \cite{foo2}:
\begin{eqnarray}
{\cal H}_{EK}=\beta \left(mV+\frac{\bm
p^2}{2m}\right)-\frac{\beta}{4m} \left\{\bm
p^2,V-1\right\}
\nonumber\\
+\frac{\beta}{2m} \left\{\bm
p^2,{\cal F}-1\right\}+\frac{\beta}{4m}\left[2
\bm\Sigma\cdot(\bm f\times\bm p)+\nabla\cdot\bm f\right]
\nonumber\\
+\frac{1}{2}\left(\bm\Sigma\cdot\bm\phi \right),
\label{eq8}\end{eqnarray}
and compare it with Eq. (\ref{eq7}) of the present work in the nonrelativistic
approximation:
\begin{eqnarray}
{\cal H}_{FW}=\beta \left(mV+\frac{\bm
p^2}{2m}\right)-\frac{\beta}{4m} \left\{\bm
p^2,V-1\right\}
\nonumber\\
+\frac{\beta}{2m} \left\{\bm p^2,{\cal F}-1\right\}+
\frac{\beta}{4m}\left[2 \bm\Sigma\cdot(\bm
f\times\bm p)+\nabla\cdot\bm f\right]
\nonumber\\
-\frac{\beta}{8m}\left[2\bm\Sigma \cdot(\bm\phi\times\bm
p)+\nabla\cdot\bm\phi\right].
\label{eq9}\end{eqnarray}
FW Hamiltonians (\ref{eq7}) and (\ref{eq9}), contrary to EK Hamiltonian
(\ref{eq8}), do
not contain the term $(\bm\Sigma\cdot\bm\phi)/2$ but contain
additional terms proportional to derivatives of $V$. These
additional terms describe both the spin-orbit and contact
interactions. To check the compatibility with \cite{Ob1,Ob2}, the
semi-relativistic transformation (with an accuracy up to $v/c$) of
Hamiltonian (\ref{eq8}) to the FW representation can be performed. With
this accuracy, transformation operator (\ref{eq4}) takes the form
\begin{equation} U_{EK\rightarrow FW}=U^{-1}_{FW\rightarrow EK}=1-\frac{\bm
p^2}{8m^2}-\frac{i(\bm\Pi\cdot\bm p)}{2m}.
\label{eq10}\end{equation} As a result, Hamiltonian (\ref{eq8}) is transformed
by operator (\ref{eq10}) to form (\ref{eq9}).

This transformation shows that the calculation fulfilled in
\cite{Ob1,Ob2} was correct. However, the Hamiltonian itself is
insufficient for an analysis of observable spin effects. One needs
to know the spin operator as well. As the Hamiltonian was obtained
in the EK representation, the spin operator (\ref{eq5}) is rather
complicated. At the same time, this operator acquires simple form
(\ref{fg}) in the FW representation. Let us stress, that only for
such a simple form of spin operator the terms of Hamiltonian may
be simply interpreted in terms of observable physical effects.
However, this is not true when a spin operator is complicated. In
particular, the term $(\bm\Sigma\cdot\bm\phi)/2$ in (\ref{eq8})
describing the dipole spin-gravity coupling disappears after the
transformation to the FW representation. Therefore, this term does
not lead to new observable effects.

%An absence of the term
%$(\bm\Sigma\cdot\bm\phi)/2$
%in Hamiltonians (\ref{eq7}) and (\ref{eq9}) is quite natural because the static

%gravitational field is considered.
%Such a field is quasi-electric, while the quantity $\bm\Sigma\cdot\bm\phi$
%characterizes the quasi-magnetic interaction.
%This interaction can exist when the field is described by a nondiagonal metric
%and created by a
%rotating body. In this case, terms proportional to
%$\bm\Sigma\cdot\bm\phi$ and $\bm\Sigma\cdot\bm f$ can appear.

\section {Equations of particle momentum and spin motion}

The problem of quantum description of particle and spin motion is very
important. However, quantum equations
of momentum and spin motion in a gravitational field were never derived.

The FW representation dramatically simplifies the derivation of quantum
equations. %of particle momentum and spin motion.
The operator equations of motion
obtained via commutators of the Hamiltonian with
the momentum and polarization operators take the form
\begin{eqnarray}
\frac{d\bm p}{dt}=i[{\cal H}_{FW},\bm p]=
-\frac{\beta}{2}\left\{\frac{m^2}{\epsilon
},\bm\phi\right\}-\frac{\beta}{2}\left\{\frac{\bm p^2}{\epsilon
},\bm f\right\}
\nonumber\\
+\frac{m}{2 \epsilon (\epsilon
+m)}\nabla\bigl(\bm{\Pi}\cdot(\bm\phi\times\bm
p)\bigr)
- \frac{1}{2\epsilon } \nabla\bigl(\bm{\Pi}\cdot(\bm
f\times\bm p)\bigr)
\label{eq11}\end{eqnarray} and
\begin{eqnarray}
\frac{d\bm\Pi}{dt}=i[{\cal H}_{FW},\bm\Pi]=
\frac{m}{\epsilon (\epsilon +m)}\bm\Sigma\times\left(\bm\phi\times\bm p\right)
\nonumber\\
-\frac{1}{\epsilon }\bm\Sigma\times\left(\bm f\times\bm p\right),
\label{eq12}\end{eqnarray}
respectively. These equations constitute our principal new result.

It is possible to prove that EK Hamiltonian (\ref{eq8}) leads to
the spin motion equation consistent with Eq. (\ref{eq12}). Within
the semi-relativistic approximation, the polarization operator in
the EK representation takes the form
$$
\bm O_{EK}=\bm\Pi+\frac{\bm p\times\bm\Sigma}{m}+\frac{\bm p\times(\bm
p\times\bm\Pi)}{2m^2}. $$
Commuting Hamiltonian (\ref{eq8}) with the polarization operator $\bm O_{EK}$
leads to the approximate equation of spin motion
\begin{eqnarray}
\frac{d\bm O_{EK}}{dt}=i[{\cal H}_{EK},\bm O_{EK}]=
\frac{\beta}{2m}\bm O_{EK}\times\left(\bm\phi\times\bm p\right)
\nonumber\\
-\frac{\beta}{m}\bm O_{EK}\times\left(\bm f\times\bm p\right)
\label{eq13}\end{eqnarray}
that agrees with Eq. (\ref{eq12}). This explicitly shows, that dipole
spin-gravity coupling cancels with the extra terms
in the spin operator in the EK representation and does not affect observable
quantities.

Let us pass to the studies of semiclassical limit of these equations.
The contribution of the lower spinor is negligible and the transition to the
semiclassical description is performed
by averaging the operators in the equations for the upper spinor \cite{JMP}. It
is usually possible to neglect the
commutators between the coordinate and momentum operators. As a result, the
operators $\bm \sigma$ and $\bm p$ should be
substituted by the corresponding classical quantities: the polarization vector
(doubled average spin), $\bm \xi$, and the momentum.
For the latter quantity, we retain the notation $\bm p$. The semiclassical
equations of motion are
\begin{eqnarray}
\frac{d\bm p}{dt}= -\frac{m^2}{\epsilon }\bm\phi-\frac{\bm
p^2}{\epsilon }\bm f+\frac{m}{2 \epsilon (\epsilon
+m)}\nabla\bigl(\bm{\xi}\cdot(\bm\phi\times\bm
p)\bigr)
\nonumber\\
- \frac{1}{2\epsilon } \nabla\bigl(\bm{\xi}\cdot(\bm
f\times\bm p)\bigr) \label{eq14}\end{eqnarray} and
\begin{equation}
\frac{d\bm\xi}{dt}= \frac{m}{\epsilon (\epsilon
+m)}\bm\xi\times\left(\bm\phi\times\bm p\right)- \frac{1}{\epsilon
}\bm\xi\times\left(\bm f\times\bm p\right),
\label{eq15}\end{equation} respectively. In Eq. (\ref{eq14}), two
latter terms describe a force dependent on the spin. This force is
similar to the electromagnetic Stern-Gerlach force (see Ref.
\cite{JMP}). Because it is weak, the approximate semiclassical
equation of particle motion takes the form
\begin{equation}
\frac{d\bm p}{dt}=
-\frac{m^2}{\epsilon }\bm\phi-\frac{\bm p^2}{\epsilon }\bm f.
\label{eq16}\end{equation}
Eq. (\ref{eq15}) can be represented as
\begin{equation}
\frac{d\bm\xi}{dt}=\bm\Omega\times\bm\xi,
\label{eq17}\end{equation}
where the angular velocity of spin rotation is given by
\begin{equation}
\bm\Omega=-\frac{m}{\epsilon
(\epsilon +m)}\left(\bm\phi\times\bm
p\right)+\frac{1}{\epsilon}\left(\bm f\times\bm p\right).
\label{eqom}\end{equation}

We can find similar equations describing a change of the direction of particle
momentum, $\bm n=\bm p/p\;$:
\begin{equation}
\frac{d\bm n}{dt}=\bm\omega\times\bm n, ~~~ \bm\omega=
\frac{m^2}{\epsilon p}\bigl( \bm\phi\times\bm
n\bigr)+\frac{p}{\epsilon}\bigl(\bm f\times\bm n\bigr).
\label{eq18}\end{equation}
Therefore, the spin rotates with respect to the momentum direction and the
angular velocity of this rotation is
\begin{equation}
\bm o=\bm\Omega-\bm\omega=-\frac{m}{p}\bigl(\bm\phi\times\bm
n\bigr). \label{eq19}\end{equation} The quantity $\bm o$ does not
depend on $\bm f$ and vanishes for massless particles. Therefore,
the gravitational field cannot change the helicity of massless
Dirac particles. The evolution of the helicity
$\zeta\equiv|\bm\xi_\| |=\bm\xi\cdot\bm n$ of massive particles is
defined by the formula
\begin{equation}
\frac{d\zeta}{dt}=(\bm
\Omega-\bm\omega)\cdot(\bm\xi_\bot\times\bm
n)=-\frac{m}{p}\left(\bm\xi_\bot\cdot\bm\phi\right),
\label{eq20}\end{equation}
where $\bm\xi_\bot=\bm\xi-\bm\xi_\|$.

\section{Particle in a spherically symmetric field}

Let us consider the interaction of particles
with a spherically symmetric gravitational field and compare the obtained
formulae with previous results.
This field is a weak limit of the Schwarzschild
one which yields
\begin{equation} V=1-\frac{GM}{r}, ~~~
W=1+\frac{GM}{r}.\label{eq21}\end{equation}
Correspondingly,
$${\cal F}=1-\frac{2GM}{r}, ~~~ \bm f=2\bm\phi=\frac{2GM}{r^3}\bm r=-2\bm g,$$
where $\bm g$ is the Newtonian acceleration.

When we neglect the terms of order of $\frac{(\bm p\cdot\nabla)(\bm p\cdot\bm g)}{\epsilon^2}$,
Hamiltonian (\ref{eq7}) takes the form
\begin{eqnarray}
{\cal H}_{FW}=\beta\epsilon -\frac{\beta}{2}\left\{\frac{\epsilon^2+\bm p^2}{\epsilon},\frac{GM}{r}\right\}
\nonumber\\
-\frac{\beta(2\epsilon+m)}{4
\epsilon (\epsilon +m)}\biggl[2\bm{\Sigma}\cdot(\bm g\times\bm p)+ \nabla\cdot\bm g\biggr].
\label{eqd}\end{eqnarray}

In this case, the operator equations of momentum and spin motion
are given by
\begin{eqnarray}
\frac{d\bm p}{dt}=
-\beta\frac{GM}{2}\left\{\frac{\epsilon^2+\bm p^2}{\epsilon},
\frac{\bm r}{r^3}\right\}
\nonumber\\
-GM\cdot\frac{2\epsilon +m}{\epsilon (\epsilon +m)}\cdot
\nabla\left(\frac{\bm{\Pi}\cdot
(\bm r\times\bm
p)}{r^3}\right),
\label{eq22}\end{eqnarray}
\begin{eqnarray}
\frac{d\bm\Pi}{dt}=-\frac{GM}{r^3}\cdot\frac{2\epsilon +m}{\epsilon
(\epsilon +m)}\bm\Sigma\times\left(\bm r\times\bm p\right).
\label{eq23}\end{eqnarray}

In Eq. (\ref{eq22}), the last term determines the gravitational SG
force. The semiclassical formula for this post-Newtonian force is
\cite{force}
\begin{eqnarray}
\bm F_{SG}=-GM\cdot\frac{2\epsilon +m}{\epsilon (\epsilon
+m)}\cdot\nabla\left(\frac{\bm{\xi}\cdot
(\bm r\times\bm
p)}{r^3}\right).
\label{eq24}\end{eqnarray}

This formula can be transformed to a more convenient form
where the quantities $\hbar$ and $c$ are kept explicit for a moment:
\begin{eqnarray}
\bm
F_{SG}=-\frac{GM\hbar}{cr^3}\cdot\frac{2\gamma+1}{\gamma+1}\left[\bm
\beta\times\bm\xi-\frac{3\bm r\bigl(\bm r\cdot[\bm
\beta\times\bm\xi]\bigr)}{r^2}\right], \label{SGF}\end{eqnarray}
$\bm \beta=\bm v/c$ and $\gamma$ is the Lorentz factor. The SG
force is of order of $\frac{\hbar\beta}{mcr}$ with respect to the
Newtonian one.

Neglecting the SG force, one get the semiclassical
equations of momentum and spin motion:
\begin{eqnarray}
\frac{d\bm p}{dt}=
\frac{\epsilon^2+\bm p^2}{\epsilon}\bm g,
\label{mm}\end{eqnarray}
\begin{eqnarray}
\frac{d\bm\xi}{dt}=\frac{2\epsilon
+m}{\epsilon (\epsilon +m)}\bm\xi\times\left(\bm g\times\bm
p\right). \label{spinm}\end{eqnarray}

The semiclassical expressions for the angular velocities of
rotation of unit momentum vector, $\bm n=\bm p/p$, and
spin are
%momentum and spin motion are
\begin{eqnarray}
\bm\omega=-\frac{\epsilon^2+\bm
p^2}{\epsilon\bm p^2}\bm g\times\bm p=\frac{GM}{r^3}\cdot
\frac{\epsilon^2+\bm p^2}{\epsilon\bm p^2}\bm l,
\label{eq25}\end{eqnarray}
\begin{eqnarray}
\bm\Omega=-\frac{2\epsilon
+m}{\epsilon (\epsilon +m)} \bm g\times\bm
p=\frac{GM}{r^3}\cdot\frac{2\epsilon +m}{\epsilon (\epsilon +m)}
\bm l, \label{eq26}\end{eqnarray} where $\bm l=\bm r\times\bm p$
is the angular moment.

Eqs. (\ref{eq25}) and (\ref{eq26}) agree with the classical
gravity. Eq. (\ref{eq25}) leads to the expression for the angle of
particle deflection by a gravitational field
\begin{equation}
\theta=\frac{2GM}{\rho}\left(2+\frac{m^2}{\bm
p^2}\right)=\frac{2GM}{\rho \bm v^2}\left(1+\bm v^2\right)
\label{eq27}\end{equation} coinciding with Eq. (13) of Problem
15.9 from Ref. \cite{AR} (see also Ref. \cite{Far}). This directly
proves the full compatibility of quantum and classical
consideration and disagrees with the results obtained in
\cite{RAINBOW}.

Eq. (\ref{eq26}) and the corresponding equation obtained in Ref.
\cite{PK} by the very different method coincide, up to the sign
due to the different definition of angular velocity \cite{foo3}.
For an immovable particle, the angular velocity of spin rotation
is described by the same formula as the de Sitter one for a classical gyroscope
\cite{DeS}. Such a similarity~\cite{KO} of classical and quantum rotators is a manifestation
of the equivalence principle (see e.g. \cite{T1,T2} and Refs. therein).
In the nonrelativistic approximation, the last term in Hamiltonian (\ref{eqd})
describing the spin-orbit and contact (Darwin) interactions coincides with the corresponding term in
Ref. \cite{DH}.

The momentum and spin rotate in the same direction.
Formula (\ref{eq19}) for the angular velocity of spin
rotation with respect to the momentum direction, defining the evolution of particle helicity, takes the form
\begin{equation}
\bm o=\bm\Omega-\bm\omega=\frac{m}{\bm p^2}\bigl(\bm g\times\bm
p\bigr). \label{rel}\end{equation}

The ratio of particle momentum and spin deflection angles ($\theta$
and $\Theta$, respectively) is constant and equal to
$$\frac{\Theta}{\theta}=\frac{(2\epsilon+m)(\epsilon-m)}{2\epsilon^2-m^2}.$$
If these angles are small, the helicity of particle, whose helicity is originally $+1$, is given by
\begin{equation}
\zeta=1-\frac{(\Theta-\theta)^2}{2}.
\label{rew}\end{equation}

Therefore, the evolution of the helicity is described by the equation
\begin{equation}
\zeta=1-\frac{\theta^2}{2(2\gamma-\gamma^{-1})^2},\label{eq28}\end{equation}
where $\gamma=\epsilon/m$ is the Lorentz factor. This equation
agrees with \cite{T1,T2} (see  Eqs. (17),(19) from \cite{T2}) obtained by the
full quantum treatment. At the same time, the expression obtained
earlier by the similar method \cite{helic94} contains the
dependence on graviton source mass $M$ and looks much more
complicated. We found that the large $M$ behavior of numerical
values as presented at Fig. 3 of that reference is at reasonable
agreement with (\ref{eq28}), while their asymptotic formula (12)
is at variance with us. Note also that in \cite{helic94} the
disagreement with the semiclassical treatment \cite{CP} was
stated, while we observe the full agreement between the
semiclassical and quantum approaches.

We may conclude for three of the most important problems formulae
(\ref{eq25}),(\ref{eq26}), and (\ref{eq28}) are in the best
agreement with previous results. We also have established a
consent between the classical and quantum theories and found the
new quantum corrections to the Newtonian force.

\section{Particle in a uniformly accelerated frame and the equivalence principle}

Consideration of the particle motion in an accelerated frame permits to relate
the gravity and acceleration. The simplest case
is the flat Minkowski spacetime in a uniformly accelerated frame
(see item (i) of Section II).
For this
problem, the exact Dirac Hamiltonian
derived by Hehl and Ni \cite{HN} is given by
\begin{equation}
{\cal H} = (1+\bm{a}\cdot\bm{r})\beta
m+\frac12\{(1+\bm{a}\cdot\bm{r}),\bm\alpha\cdot\bm p\},
\label{eq29}\end{equation}
where $\bm{a}$ is the particle acceleration. In this case, the metric tensor
corresponds to the choice (\ref{acf}).
Metric (\ref{acf}) corresponds to the following form of FW Hamiltonian
(\ref{eq7}):
\begin{equation}\begin{array}{c}
{\cal H}_{FW}=\beta\left(\epsilon
+\frac{1}{2}\{\epsilon,\bm{a}\cdot\bm{r}\}\right)
+\frac{\bm{\Pi}\cdot(\bm a\times\bm p)}{2(\epsilon +m)},
\end{array}\label{eq30}\end{equation}
where $\epsilon =\sqrt{m^2+\bm p^2}$. The contact (Darwin)
interaction does not appear because the effective field $\bm a$ is
uniform. Equation (\ref{eq30}) shows that the particle energy is
multiplied by the factor $V$ except for the last term that is of a
purely quantum origin. An appearance of this term describing the
inertial spin-orbit coupling has been discovered by Hehl and Ni
\cite{HN}. In the present work, generalizing the result of
this reference, %unlike to this reference,
the relativistic expression for the Hamiltonian has been derived.
%,which
This expression happens to agree with the nonrelativistic
ones from \cite{HN,Huang}.

Eq. (\ref{eq30}) for the Hamiltonian of relativistic particle
in a uniformly accelerated frame agrees with the corresponding
nonrelativistic expressions from \cite{HN,Huang}.

The equations of particle and spin motion are given by
\begin{equation}
\frac{d\bm p}{dt}= -\beta\epsilon \bm a, ~~~ \frac{d\bm\Pi}{dt}=-
\frac{\bm\Sigma\times(\bm a\times\bm p)}{\epsilon +m}.
\label{eq31}\end{equation}
In the uniformly accelerated frame, the SG force does not exist.

The semiclassical transition brings Eq. (\ref{eq31}) to the form
\begin{equation}
\frac{d\bm p}{dt}= -\epsilon \bm a, ~~~ \frac{d\bm\xi}{dt}=-
\frac{\bm\xi\times(\bm a\times\bm p)}{\epsilon +m}.
\label{eq32}\end{equation}

The angular velocities of rotation of unit momentum vector
and spin are equal to
\begin{equation}
\bm\omega=\frac{\epsilon}{\bm
p^2}\bigl(\bm a\times\bm p\bigr), ~~~ \bm\Omega=\frac{\bm
a\times\bm p}{\epsilon +m}. \label{eq33}\end{equation}

The relative angular velocity defining the helicity evolution is given by
\begin{equation}
\bm
o=\bm\Omega-\bm\omega=-\frac{m}{\bm p^2}\bigl(\bm a\times\bm
p\bigr). \label{eq34}\end{equation}

When $\bm a=-\bm g$, values of $\bm o$ in Eqs. (\ref{eq34}) and (\ref{rel}) are the same. It is the manifestation
of the equivalence principle which was discussed with respect to helicity evolution in  \cite{T1,T2}.

At the same time, the manifestation of the equivalence principle for  the spin
rotation is not so trivial.  In particular, the spin of
nonrelativistic particles in the spherically symmetric
gravitational field rotates three times more rapidly in comparison to the accelerated frame.

To trace the origin of this difference, let us compare the rotation of the momentum direction in these cases.
Although it is the same in the nonrelativistic limit, the expressions for the relativistic particles differ.
To understand this from the point of view of equivalence principle, the approach of \cite{T1,T2} is convenient.
Let us consider \cite{T1,T2} the matrix element ${\cal M}$ of particle scattering in the external gravitational field
\begin{eqnarray}
\label{0g}
{\cal M}=\frac{1}{2}
\langle p{'}| T^{\mu \nu} |p \rangle h_{\mu \nu}(q),~~~ q=p-p{'},
\label{A1}
\end{eqnarray}
where  $T^{\mu \nu}$ is the Belinfante energy-momentum tensor  and $h_{\mu \nu} $
is a Fourier component of a deviation of the metric tensor from its Minkowski value.
The particle momentum evolution is fully determined by the forward matrix element fixed by the momentum conservation
\begin{eqnarray}
\langle p| T^{\mu \nu} |p\rangle=2 p^\mu p^\nu.
\label{e2}
\end{eqnarray}
The matrix element for the particle at rest takes the form
\begin{eqnarray}
{\cal M} =  m^2 h_{0 0}(q).
\label{A2}
\end{eqnarray}
Coincidence of  the $(00)-$components of the metric in the gravitational field and accelerated frame
proves the equivalence principle, appearing in such an approach as a low-energy theorem
rather than postulate.

At the same time, for the moving particle, the space components of metric $h_{zz}=h_{xx}=h_{yy}=h_{00}$ (see e.g. \cite{DeS}) also contribute.
As a result, the matrix elements in the gravitational field (${\cal M}_g$) and in the accelerated frame ($ {\cal M}_a$)
differ by the obvious kinematical factor:
\begin{eqnarray}
{\cal M}_g= (\epsilon^2+\bm p^2) h_{0 0}(q),~~~ {\cal M}_a=   \epsilon^2 h_{0 0}(q).
\label{A3}
\end{eqnarray}
The ratio of these matrix elements
\begin{eqnarray}
R=\frac{\epsilon^2+\bm p^2} {\epsilon^2}
\label{ratio}
\end{eqnarray}
is exactly equal to the ratio of the r.h.s. of the
equations of particle momentum motion, and, consequently, to the ratio of the
angular velocities of rotation of their directions. It is now clearly seen
that this difference is a direct kinematical consequence of the equivalence principle.

Note that general expression (\ref{eq20}) for the helicity evolution is insensitive to the space components of the metric
which is the entire origin of the kinematical factor differing in the gravitational filed and accelerating frame.
This provides the additional argument for the simple form of the equivalence principle when helicity
is considered. Namely, the helicity evolution in any static gravitational field and corresponding accelerating frame merely coincides.

Let us consider the effect of the mentioned kinematical factor for the spin motion.
 Eqs. (\ref{eq25}) and (\ref{eq33}) describing the angular velocity of momentum motion can be written in the form
\begin{eqnarray}
\bm\omega_g=-\left[\frac{m}{\bm p^2}+\frac{2\epsilon+m}{\epsilon(\epsilon+m)}\right]
\bigl(\bm g\times\bm
p\bigr)
\nonumber\\
=-\bf o-\frac{2\epsilon+m}{\epsilon(\epsilon+m)}
\bigl(\bm g\times\bm p\bigr)
\label{eq36}\end{eqnarray}
for the spherically symmetric gravitational field and
\begin{equation}
\bm\omega_a=\left(\frac{m}{\bm p^2}+\frac{1}{\epsilon+m}\right)
\bigl(\bm a\times\bm
p\bigr)
%\nonumber\\
=-\bf o+\frac{1}{\epsilon+m}
\bigl(\bm a\times\bm p\bigr) \label{eq37}\end{equation}
for the uniformly accelerated frame.
Here the relative angular velocity $\bm o$, common for two cases, is extracted.
The remaining terms in the r.h.s. are just the angular velocities of spin rotation.
The current derivation explicitly shows that their difference is the
consequence of the equivalence principle and kinematical factors in (\ref{A3}).

It is interesting that in the nonrelativistic limit both $\bm \omega$ and $\bm o$
diverge as $1/\bm p^2$.
Their finite differences in Eqs. (\ref{eq36}) and (\ref{eq37}) provide the nonrelativistic limit of the angular velocities
of spin rotation.
In this limit  the  momentum rotation in the gravitational field and accelerating
frame coincides, as it is seen from (\ref{ratio}). However, the mentioned divergence
of the angular velocities ``compensates" the infinitesimal deviation
of $R$ from unity. In more detail, 
one is dealing with the low $\bm p^2$ expansion of two expressions 
\begin{eqnarray}
\bm\omega_g=-\frac{\epsilon^2+\bm
p^2}{\epsilon\bm p^2}\bigl(\bm g\times\bm p\bigr) \approx -\left(\frac{m}{\bm p^2} + \frac{3}{2m}\right)\bigl(\bm g\times\bm p\bigr)
\label{eqap2}\end{eqnarray}
and 
\begin{equation}
\bm\omega_a=\frac{\epsilon}{\bm
p^2}\bigl(\bm a\times\bm p\bigr) \approx   \left(\frac{m}{\bm p^2} + \frac{1}{2m}\right)\bigl(\bm a\times\bm p\bigr).
\label{eqap1}\end{equation}
While the l.h.s of these expressions has the same nonrelativistic limit, the mentioned effect 
provides the ratio 3 for finite terms in r.h.s.

For completeness let us also consider 
the operator equation for the particle acceleration:
\begin{equation}
\ddot{\bm{r}}=-\bigl[{\cal H},[{\cal H},\bm
r]\bigr]=-\frac{1}{2}\left\{(1+\bm a\cdot\bm r),\left[\bm
a-\frac{2\bm p (\bm a\cdot\bm p)}{\epsilon^2}\right]\right\}.
\label{acso}\end{equation}

In this equation, small terms depending on the spin matrix
are omitted. In the semiclassical approximation,
\begin{eqnarray}
\ddot{\bm{r}}=-(1+\bm a\cdot\bm r)\left[\bm a-\frac{2\bm p (\bm
a\cdot\bm p)}{\epsilon^2}\right] \nonumber\\
=-(1+\bm a\cdot\bm
r)\bm a-\frac{2\bm v (\bm a\cdot\bm v)}{1+\bm a\cdot\bm r}.
\label{acse}\end{eqnarray}
After substitution of standard nonrelativistic expressions for $\bm r$ and  $ \bm v$ we reach
the full agreement with the approximate result of Huang and Ni
(\cite{HuangNi}, Eq. (82)).

\section {Discussion and summary}

We showed that the elegant exact EK transformation \cite{Ob1,Ob2}
does not provide a simple form for dynamical operators and
therefore does not allow for a straightforward derivation of
quantum and semiclassical equations of motion. We constructed the
FW transformation leading to simple dynamical operators and
derived the quantum (which is our main new result) and
semiclassical equations of momentum and spin motion. For the case
of weak spherically symmetric field the semiclassical limit
reproduces all the known results for the momentum, spin and helicity
evolution and resolve the existing contradictions. 
The new quantum corrections provide, in
particular, the post-Newtonian gravitational SG force. We found
that semiclassical equations are in full agreement with classical
gravity.

We checked that the derived equations of motion are compatible with those
obtained from Hamiltonian of
\cite{Ob1,Ob2} and the respective (complicated) dynamical operators.
However, the difference %discrepancy
between the FW and EK representations means that the  physical
interpretation of the approach \cite{Ob1,Ob2} should be made carefully. 
Say, the term in Hamiltonian \cite{Ob1,Ob2} describing the
dipole spin-gravity coupling does not appear in the FW
Hamiltonian. As soon as physical effects are dependent on both the
Hamiltonian and dynamical operators, the correspondent term in the
EK representation is cancelled, when complicated spin operator
(\ref{eq5}) is used. Consequently, there is no reason for
the precession of spin of particles being at rest, which is
explicitly seen from Eq. (\ref{eq12}).
%As a result, quantum and classical equations of spin motion are in full
%agreement.
%This is a necessary condition for the validity of post-Newtonian
%equivalence principle (see, e.g., Ref. \cite{T}).

The equivalence principle, understood as minimal coupling of fermions to gravity (\ref{eqin}),(\ref{A1}) is always valid.
However, its specific manifestations depend
on the observable. From this
point of view, the simplest observable for Dirac particle is {\it helicity}.
The helicity evolution in the gravitational field and accelerated frame
is {\it the same}.
The manifestation of the equivalence principle for momentum and spin motion in these two cases
is affected by kinematical corrections due to the space
components of metric tensor. In particular, this leads to the enhancement by the factor 3 of the
frequency of spin precession in the gravitational field with respect to the accelerating frame.

\section*{Acknowledgements}

We are indebted to F.W.~Hehl, I.B.~Khriplovich,  C.~Kiefer, L.~Lusanna and Yu.N.~Obukhov for discussions and correspondence.
We acknowledge a financial support by the BRFFR
(Grant $\Phi$03-242) and RFBR (Grant 03-02-16816).

\end{document}